\documentstyle[11pt,paspconf,epsfig]{article}

\def\gsim{\lower.73ex\hbox{$\sim$}\llap{\raise.4ex\hbox{$>$}}}
\def\lsim{\lower.73ex\hbox{$\sim$}\llap{\raise.4ex\hbox{$<$}}}

\begin{document}

\title{Galaxy Counts, Sizes, Colours and Redshifts in the Hubble Deep Field}

\author{T. Shanks, N. Metcalfe, R. Fong,  H.J. McCracken,} 
\affil{Department of Physics, University of Durham, South Road,\\
Durham DH1 3LE, UK}

\author{ A. Campos,}
\affil{ Observatorio Astronomico Nacional, 28800 Alcal de Henares, Madrid,
Spain}

\author{J.P. Gardner}
\affil{Laboratory for Astronomy and Solar Physics, Code 681, Goddard Space
Flight Center, Greenbelt MD20771, USA}



\begin{abstract}
We compare the galaxy evolution models of Bruzual \& Charlot (1993)
with faint galaxy count, size and colour data from the Hubble and Herschel
Deep Fields (Metcalfe et al 1996). For $q_o=0.05$, we find that models
where the SFR increases  exponentially out to $z>2$ are consistent with all
of the observational data.  For  $q_o=0.5$, such models require an extra
population of galaxies which are only seen at high redshift and then
rapidly fade or disappear. We find that, whatever the cosmology, the
redshift of the faint blue galaxies and hence the epoch of galaxy formation
is likely to lie at $z\gsim2$. We find no implied peak in the SFR at
$z\approx1$ and we suggest that the  reasons for this  contradiction with
the results of Madau et al (1996) include differences in faint galaxy
photometry,  in the treatment of  spiral dust and  in the local galaxy
count normalisation. \end{abstract}




\section{Hubble and Herschel Deep Field Data}

We have  measured galaxy counts, sizes and colours in the Hubble Deep Field
(HDF, Williams et al 1996) to $U=27.^m5$ $B=29^m$, $R=28.^m5$ and
$I=28^m$  (Metcalfe et al 1996).  We  have further measured galaxy counts
and colours in a 50hr B CCD exposure to $B=28.^m2$ using the William Herschel
Telescope (WHT) in the $7'\times7'$ Herschel Deep Field. Within this field we
have also obtained a 30hr IRCAM3 exposure with  UKIRT to $K=22.^m75$ in a
$1.'3\times1.'3$ area.

We show in Fig. 1 our derived  B, I and K galaxy number counts, together
with those from  other work.  The HST B counts extend one magnitude deeper
than the WHT data and they are in good agreement where they overlap.  The
HST  I counts extend about $10\times$ deeper than previous  data, because of
the higher HST resolution and the fainter background sky.  The K counts  now
also  appear to be well defined in the range $15^m<K<24^m$.

\section{Galaxy Count Models}
\begin{figure}
\centerline{\epsfig{figure=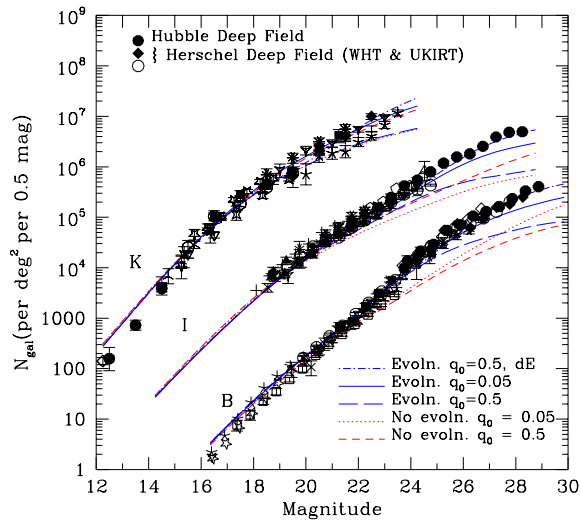, height=3in}}
\caption{B, I($\times10$) and K($\times100$) galaxy counts from  the Hubble
and Herschel Deep Fields  compared to previous counts and  various models. 
Using Bruzual \& Charlot(1993) evolution models with an exponentially
increasing SFR, we adopt a dwarf dominated IMF (x=3) with  $\tau$=2.5~Gyr, for
E/SO/Sab galaxies and a Salpeter IMF (x=1.35) with  $\tau$=9~Gyr, for
Sbc/Scd/Sdm  galaxies. Spiral dust absorption is taken to be $A_B=0.^m3$ at 
$z=0$. The models also include  Lyman $\alpha$ absorption (Madau 1995). The
$q_o=0.05$ models give good fits to both the highly evolved B,I counts  and
the more slowly evolving K counts. The $q_o=0.5$ models require the
presence of an extra galaxy (disappearing dwarf, dE) population at high
redshift to fit the faint B and I counts.}   \end{figure}

As noted previously (Shanks 1990, Koo \& Kron 1992, Metcalfe et al 
1991,1995), if the B count models are normalised at $B\approx18^m$ rather
than $B\approx15^m$ then non-evolving models give a reasonable representation
of the B band counts and redshift distributions in the range $18^m<B<22.^m5$.
This high normalisation has recently received new support from HST galaxy
counts subdivided by morphology where non-evolving models with the high
normalisation give an excellent fit to both spiral and early type galaxy
counts with $17^m<I<22^m$ (Glazebrook et al 1995a, Driver et al 1995). The B
galaxy count then shows evidence for strong evolution in the range
$23^m<B<29^m$ (see Fig. 1). Also the high normalisation allows non-evolving
models with $0.05<q_o<0.5$ to fit the less steep K counts to $K\approx24^m$. 

Simple evolutionary models where galaxy  star-formation rates rise
exponentially with look-back time, are known to fit the B counts in
the range $18^m<B<25^m$ (Koo \& Kron 1992, Metcalfe et al 1991,1995). But
previous faint galaxy redshift surveys at $B<24^m$ presented a problem for
such models, as they  predict a high  redshift tail which was unobserved in
these early surveys (Glazebrook et al 1995b). However, Cowie et al
(1995,1996) have recently used the Keck 10m telescope to make a new
$22.^m5<B<24^m$ galaxy redshift survey with $>$80\% completeness and have now
detected such a high redshift galaxy component, supporting the viability of
these models. 

Furthermore, spiral evolution models with exponentially increasing
star-formation rates of time scale $\tau\approx$9~Gyr, are now thought to be
able to sustain large amounts of B band evolution (Bruzual \& Charlot 1993,
Campos \& Shanks 1997).  These spiral  models also have the advantage that
evolving the steeper luminosity function of late-type galaxies (Shanks 1990,
Metcalfe et al 1995) makes it easier to increase the B number counts using
lower redshift galaxies, particularly if they contain even small amounts of
dust (Wang 1991, Gronwall \& Koo 1995, Campos \& Shanks 1997).

In the low $q_o$ case, this spiral dominated   model then produces a
reasonable fit to the counts to $B\approx27^m$ and $I\approx26^m$ and 
$K\approx24^m$ (see Fig. 1). Although the model underestimates the optical
counts at fainter magnitudes,  this discrepancy is probably still within the
combined data and  model uncertainties. In the $q_o=0.5$ case, the spiral
luminosity evolution model only fits the optical data  to $B\approx25^m$ and
$I\approx23.^m5$ and then more seriously underestimates the counts at fainter
magnitudes.  Thus, the HST data confirms the previous ground-based result
(Yoshii \& Takahara 1988, Koo \& Kron 1992, Metcalfe et al 1995) that if
$q_o=0.5$, then there is not enough spatial volume at high redshifts to allow
simple luminosity evolution models to fit the high galaxy counts at $B>25^m$.
To improve the fit of the  $q_o=0.5$ model, we consider a  model with an extra
population of high redshift galaxies which have a constant star-formation
rate from the formation epoch till z=1. The Bruzual \& Charlot model
shows that at  $z\approx 1$ the galaxy then rapidly fades by $5^m$ in B to 
form a red  (dE) galaxy by the present day. This model is in the spirit of
previously proposed `disappearing dwarf' count models (Babul \& Rees 1992)
and it gives  a good fit to the faint B,I,K counts. 

Finally, Metcalfe et al  (1996, Fig. 2) have  further shown that both the
$q_o=0.05$ and $q_o=0.5$ models also  fit  the new Keck $22.^m5<B<24^m$ n(z) 
data of Cowie et al (1996). The models also fit the apparently unevolved Keck
$18^m<K<19^m$ n(z) of Cowie et al by virtue of our assumed x=3 IMF for
early-type galaxies which reduces their essentially passive K evolution to
acceptable limits.

\section{Galaxy Sizes}

\begin{figure}
\centerline{\epsfig{figure=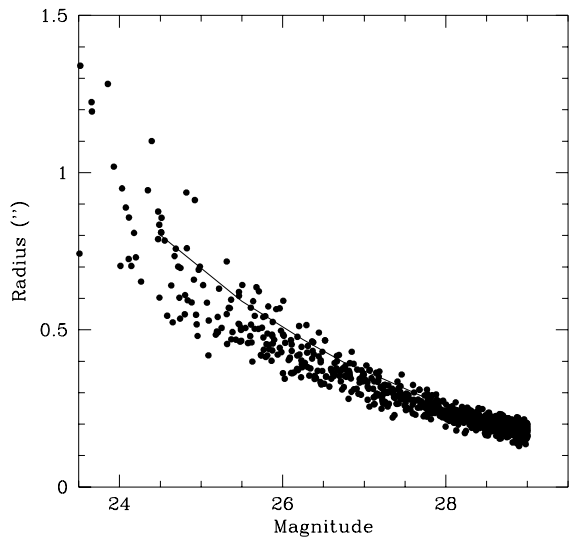,
height=3in}} \caption{ The isophotal ($\mu_B$=28mag arcsec${}^{-2}$) radius
versus B magnitude relation for HDF data (dots) shows reasonable agreement
with the average result ($q_o=0.05$) from a PPP simulation of an HDF frame 
(solid line) using our evolutionary models. } 
\end{figure}

It has been claimed that  angular  sizes of  faint HDF  galaxies are small
compared to local galaxies (Roche et al 1998 and references therein). It
is therefore interesting to see if the angular size data can reject the above
models. So, following Metcalfe et al(1995), we adapted the PPP CCD simulation
code (Yee \& Green 1987) to simulate an HDF frame. We assumed Freeman's(1970)
law to relate disk size to absolute magnitude at z=0 and assumed that
luminosity  evolution only affects disk surface brightness and not size. For
bulges, we similarly used an  $r^{1/4}$ law and the diameter-magnitude
relations of Sandage \& Perelemuter (1990). In the $q_0=0.5$ case we made the
same `bulge' assumptions for the `disappearing dwarf' population. We then
applied our photometry software to determine the B magnitude-angular size
relation to $B=28^m$ for both data and simulation. Fig. 2 shows that although
the HDF galaxy isophotal radii are small and only of order $0.''2$, the
$q_0=0.05$ model also predicts similarly small sizes for the galaxies. The
same is true for the $q_0=0.5$ model. A small size is predicted because we
are looking  down the luminosity function at $B=28^m$ to galaxies with low
intrinsic luminosities and hence small sizes.

\section{Model Constraints from Galaxy Colours and Redshifts}

We next test the  models against the faint galaxy colour distributions
in the HDF. The presence of broad  features  in galaxy spectra allows tests to
be made of the predicted galaxy redshifts.  Here we simply compare  our
evolutionary model predictions to the  observations of faint galaxy colours. 
Metcalfe et al (1996, Fig. 3) showed  that our  predicted U-B:B-I model
tracks compare well with the observed colours for $B<27.^m5$ galaxies. The
redshifting of the Lyman $\alpha$ forest/break (Madau, 1995) through the U
band causes the model U-B colours to move sharply redwards and the same
effect is clearly seen in the data. Metcalfe et al found that the proportion
of galaxies with U-B$>$0 colours corresponding to $z>2$ is approximately half
of the total at $27^m<B<28^m$, indicating that the apparent redshift
distribution may peak at z$\approx$2, in good agreement with both the
$q_o=0.05$ evolutionary model and the $q_o=0.5$, disappearing dwarf model.

These conclusions are confirmed by consideration of the B-R:R-I
colour-colour plot in Fig. 3 which shows  excellent agreement between the
predicted tracks of the galaxy types with redshift and the $R<27^m$ HDF data.
The highest density peak in this plot also corresponds to the predicted galaxy
colours for  z$\approx$2. Fig. 3 also shows that there is consistency between
the galaxies predicted to have $z>2$ on the basis of both their U-B$>$0
colour and their position on the B-R:R-I plane. Metcalfe et al (1996) have
further shown that the models also give excellent agreement with the observed
distribution of  galaxy numbers across the B-R:R-I diagram. Our conclusion
is that simple models with an exponentially increasing SFR to $z>2$ can fit
the colours and redshifts of the faintest HDF galaxies as well as their
angular sizes and counts.

\begin{figure}
\centerline{\epsfig{figure=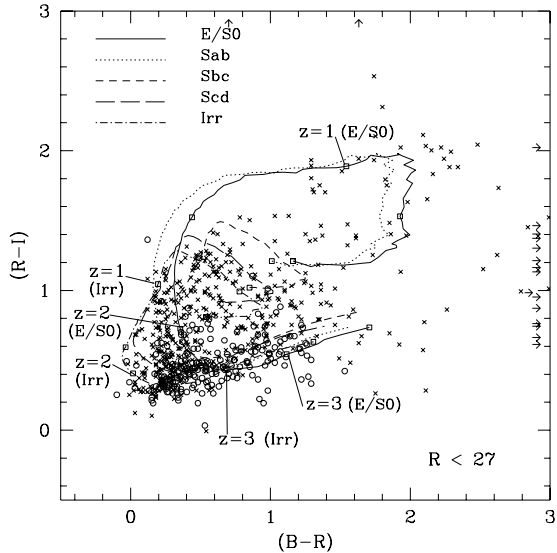,height=3in}}
\caption{The HDF B-R:R-I galaxy colours compared  to the $q_o=0.05$
evolutionary model colours. Galaxies with B-R$\approx0.^m3$ and
R-I$\approx0.^m5$ are predicted to have z$\approx$2. The circles indicate
galaxies which are predicted  to have $z>2$ on the basis of their  U-B$>$0
colour and show that they lie in a  position which consistent with $z>2$ in
this independent  plane and which is distinct from the galaxies with
U-B$<$0.}  
 
\end{figure}

\section{Comparison with the Results of Madau et al (1996).}

Madau et al (1996) have taken a different approach and, following Cowie
(1988) and  Lilly et al (1995,1996),  have used the fact that the galaxy UV
flux density is proportional to the SFR density to  obtain estimates of the
SFR history of the Universe. They use the CFRS spiral evolving LF of Lilly et
al (1995)  to determine  the 2800\AA~ flux density in the range $0<z<1$ and 
the HDF U(F300W) and B dropouts to obtain the 1620\AA~ flux density in the
ranges  $2<z<3.75$ and $3.5<z<4.5$. Fig. 4 shows their result which implies a
peak in the SFR at $z\approx1$. Also drawn on Fig. 4 is our $\tau$=9~Gyr
exponentially increasing SFR for spirals which we have shown fits a variety
of faint galaxy count, size, colour and redshift data over a wide range of
passbands. Clearly the two SFR histories are in contradiction, with the SFR
evolution rate of Madau et al increasing markedly faster than our exponential
in the $0<z<1$ range and then quickly decreasing below our rate at $z>1$. In
considering  possible reasons for the differences in the $z<1$ range, the
first point to be checked is whether our exponential models fit the
evolving luminosity functions determined from the CFRS data. Fig. 5 shows a
comparison between our predicted luminosity functions for both spirals and
ellipticals, assuming $q_o=0.05$, and the CFRS results in the redshift
intervals shown out to z=1.3. It can be seen that the agreement is generally
excellent. In Fig. 4, the poor agreement in the $0.75<z<1$ bin   may be
because Madau et al have used an LF extrapolated UV density  for the   point
at $0.75<z<1$(L. Tresse priv. comm.); if the unextrapolated luminosity
density is used then the result would be a factor of $\approx$2 lower and in
much better agreement with our result (see Lilly et al 1996 Fig. 1). At lower
redshifts the major disagreement arises at z=0 where the SFR of Gallego et
al(1995) lies a factor of 2  below our exponential model. This problem may be
related to the bright B count normalisation issue, discussed above. Supported
by the results of  Glazebrook et al (1995a) and Driver et al (1995), we
explicitly ignore the low galaxy count at $B<17^m$ as being possibly
contaminated by the effects of local large-scale structure. A further test 
should soon be available from the $B<19.^m5$ 2dF galaxy redshift survey
results of Maddox et al (1998, this vol.) where the luminosity function at low
$z$ may be expected to move in the density rather than the luminosity
direction if this large-scale structure hypothesis is correct. 

\begin{figure}
\centerline{\epsfig{figure=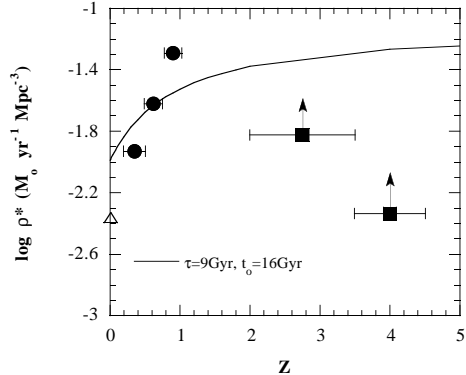,
height=2in}} \caption{The SFR-redshift plot of Madau et al (1996) 
compared to our $\tau=9$~Gyr exponentially increasing SFR for spiral
galaxies.}  \end{figure}

In the $2<z<2.75$ range, the reasons for the factor of 3 discrepancy seen in
Fig. 4 are also clear. First, if we repeat the Madau et al measurement in
the HDF, then we obtain a factor of 1.56$\times$ bigger luminosity density
at 1620\AA~ for $z>2$ UV drop-out galaxies than these authors, mostly due to
the fact that we measure brighter magnitudes for individual galaxies.
Second, if Madau et al were to include our $A_B(z=0)=0.^m3$ dust absorption
for spirals, then this would increase their 1620\AA~ luminosity density by
another factor of 1.63. Combined, these two effects give a factor of
$2.54\times$ overall which is within 20\% of the above factor of 3. In the
$3.5<z<4.5$ bin the same two effects apply,  with the increasingly serious
effects of luminosity function incompleteness also applying at the larger
distance. Note that apparently decreasing numbers of galaxies are 
seen at $z>2$ in our models due to the twin effects of distance and dust (see
Metcalfe et al 1996, Fig. 4), even though in these models the galaxy
luminosity continues to increase exponentially to $z>6$!

\begin{figure}
\centerline{\epsfig{figure=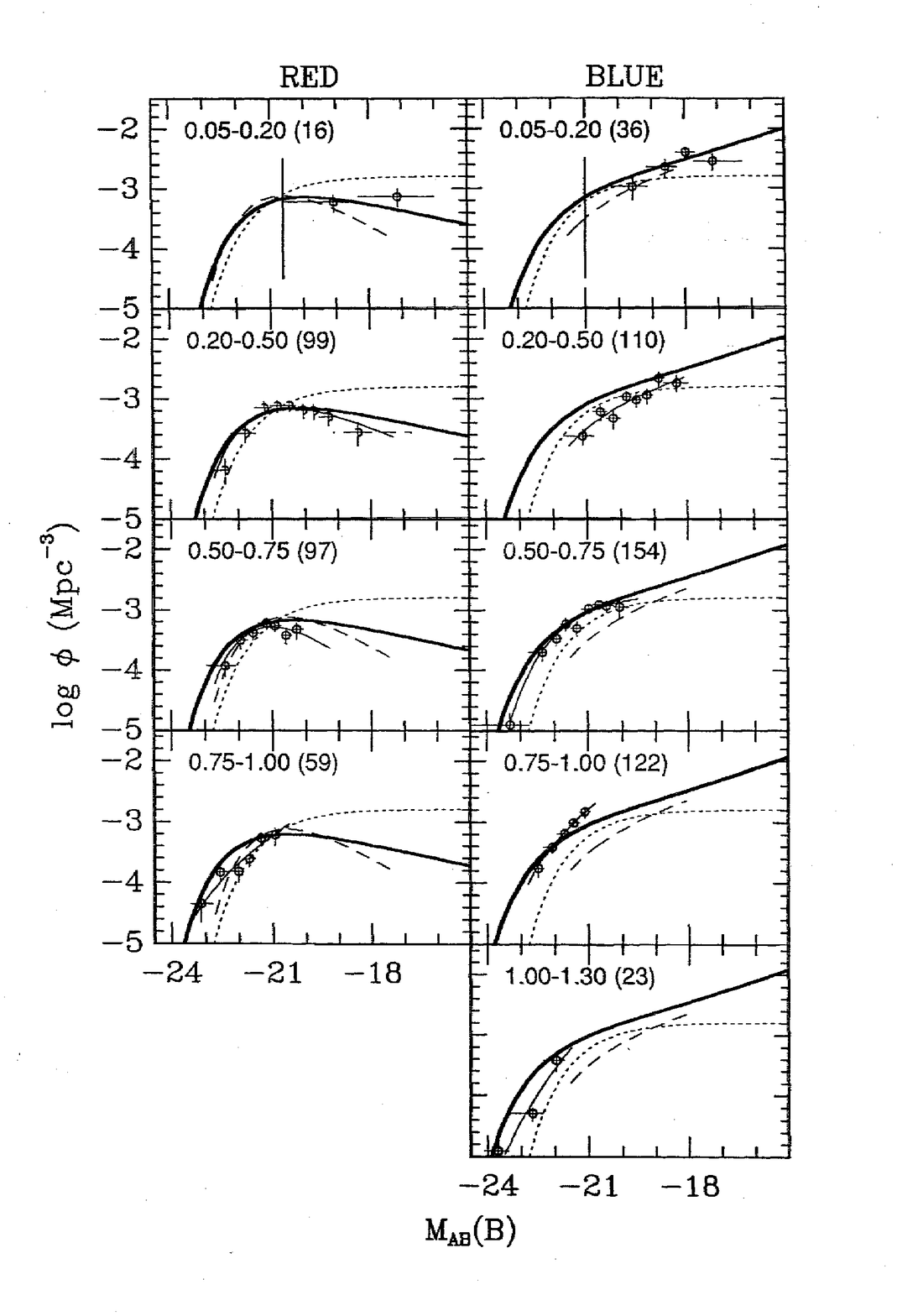,
height=4.5in}} \caption{A comparison of our model predictions for the
luminosity functions of both early and late-type galaxies (heavy, solid
lines) shows good agreement with the CFRS LF data of Lilly et al (1995). }  
\end{figure}

\section{Conclusions.}

We have shown that  simple Bruzual and Charlot models with an exponentially
increasing SFR, fit the HDF faint galaxy count, size,  colour and redshift
data over a wide range of passbands. In the $q_o=0.05$ case, the galaxy count
fit is good to $B=27^m$ whereas in the $q_o=0.5$ case, a `disappearing dwarf'
population is required to improve the fit at $B>25^m$. The HDF galaxy colours
imply that `faint blue galaxies' have $z\approx2$; if these are primordial
galaxies as previously suggested by Cowie et al (1988), then this implies the
epoch of galaxy formation is at $z\gsim 2$. The models suggest an
exponentially increasing SFR for spirals with $\tau=9$~Gyr. Our conclusion is
therefore in disagreement with the result of Madau et al (1996) and we have
indicated that the different treatments of dust and the local galaxy density,
together with photometry errors are among the factors which contribute to the
discrepancy.





%
%

\begin{references}

\reference  Babul, A. \&  Rees,M.J., 1992,  \mnras,   255,   346   

\reference  Bruzual, A.G. \&  Charlot, S. 1993, \apj,   405,   538   

\reference  Campos, A. \&  Shanks, T. 1997, \mnras,  291, 383   

\reference  Cowie, L.L. 1988 in The Post Recombination Universe eds. Kaiser,
            N. \& Lasenby, A., Dordrecht:Kluwer, 1

\reference  Cowie, L.L.,  Hu, E.M. \&  Songaila, A. 1995,
            Nature,   377,   603  

\reference  Cowie, L.L.,   Songaila, A.,  Hu, E.M. \&  Cohen, J.G. 1996,
            \aj,  112, 839   
      
\reference  Driver, S.P.,   Windhorst, R.A., Ostrander, E.J., Keel,  W.C.,
            Griffiths, R.E. \&  Ratnatunga, K.U. 1995,  \aj,  449, L23  

\reference Freeman, K.C. 1970, \apj,160, 811

\reference Gallego,J., Zamorano,J., Aragon-Salamanca,A.,\& Rego M.
           1995, \apj, 455, L1

\reference  Glazebrook, K.,  Ellis, R.S.,  Santiago, B.  \&  Griffiths, R.E.
            1995a, \mnras,  275,   L19  

\reference  Glazebrook, K.,  Ellis, R.S.,   Colless, M.M.,  Broadhurst, T.J.,
            Allington-Smith, J.R. \& Tanvir,  N.R. 1995b,  \mnras,  273,  
            157   

\reference  Gronwall, C. \&  Koo, D.C. 1995, \apj,  440,   L1   

\reference  Koo, D.C. \&  Kron, R.G. 1992, \astap,  30,   613  

\reference  Lilly, S.J., Tresse, L., Hammer, F., Crampton, D., and Le Fevre,
            O. 1995, \apj, 455, 108

\reference  Lilly, S.J., Le Fevre, O., Hammer, F. and  Crampton, D. 1996 
            \apj, 460, L1

\reference  Madau, P.  1995, \apj,    441,   18   

\reference  Madau, P., Ferguson, H., Dickinson, M., Giavilisco, M.,
            Steidel, C., Fruchter, A. 1996 \mnras, 283, 1388

\reference  Metcalfe, N., Shanks,  T.,  Fong,  R. \&  Jones, L.R. 1991, \mnras, 
            249, 481   

\reference  Metcalfe, N., Shanks,  T.,  Fong,  R. \&  Roche, N. 1995, \mnras,
            273, 257 

\reference  Metcalfe, N., Shanks,  T.,Campos, A., Gardner, J.P. \&  Fong,  R.
            1996, Nature, 210, 10

\reference Roche, N., Ratnatunga, K, Griffiths, R.E., Im, M., \& Naim, A.
           1998,  \mnras, 293, 157

\reference Sandage A., \& Perelemuter, J.-M. 1990, ApJ, 361, 1

\reference Shanks, T. 1990, in  The Galactic and Extragalactic
           Background Radiations,  S. Bowyer \& C. Leinert, Dordrecht:Kluwer,
           269

\reference Wang, B. 1991, \apj, 383, L37    

\reference Williams, R.E. {\it et al} 1996,  \aj, 112, 1335

\reference Yee, H.K.C. \& Green, R.F. 1987, \apj, 319, 218

\reference Yoshii,  Y.  \&  Takahara,  F. 1988, \apj, 326, 1

\end {references}

\end{document}